\documentstyle[12pt,aaspp4,epsf]{article}

\def\ltsim{ \,{}^<_\sim\, }
\def\gtsim{ \,{}^>_\sim\, }

\def\etal{et al.}
\def\hst{{\it HST}}
\def\ic{IC~4051}

\lefthead{Woodworth \etal }
\righthead{{\ic} Globular Clusters}

\begin{document}

\title{The Globular Cluster Systems in the Coma Ellipticals. III: 
The Unique Case of {\ic}}

\author{Sean C.~Woodworth and William E.~Harris}
\affil{Department of Physics \& Astronomy, 
McMaster University, Hamilton, Ontario L8S 4M1
\\Electronic mail: woodwrth, harris@physics.mcmaster.ca}


\begin{abstract}
IC 4051 is a giant E galaxy on the outskirts of the Coma cluster core.
Using archival \hst\ WFPC2 data, we derive
the metallicity distribution, luminosity
function, and spatial structure of its globular cluster system (GCS).
The metallicity distribution derived from the $(V-I)$
colors has a mean $\langle$Fe/H$\rangle \simeq -0.3$, 
a near-complete lack of
metal-poor clusters, and only a small metallicity gradient with radius.
We tentatively suggest that the GCS has two 
roughly equal metallicity subcomponents,
one centered at [Fe/H] $\sim 0.0$ and the second at [Fe/H] $\sim -1.0$,
although their identification is blurred by the photometric
uncertainties.
The luminosity distribution (GCLF) has the standard Gaussian-like
form observed in all other giant E galaxies, with a peak (turnover)
at $V^0 = 27.8$, consistent with a Coma distance of 100 Mpc.
The radial profiles of both the GCS and the halo light
show an unusually steep falloff which may indicate that the halo
of this galaxy has been tidally truncated.
Lastly, the specific frequency
of the GCS is remarkably large:  we find $S_N = 11 \pm 2$, at a level
which rivals M87 and most others in the central cD-type category, even
though IC 4051 is not a cD or brightest cluster elliptical.
This galaxy exhibits a combination of GCS characteristics
found nowhere else.  A formation model consistent
with most of the observations would be that this galaxy was subjected to 
removal of a large fraction of its protogalactic gas shortly after
its main phase of globular cluster formation, probably by its first
passage through the Coma core.  Since then, no significant additions
due to accretions or mergers have taken place, in strong contrast
to the central Coma galaxy NGC 4874.

\end{abstract}


\clearpage

\section{INTRODUCTION}

The Coma Cluster ($d \sim 100$ Mpc), as a rich Abell cluster, is the 
host environment for a huge range of E/S0 galaxies.  Its two
central supergiants, NGC 4874 and 4889, are among the very most luminous
galaxies known, and the cluster has many other large ellipticals
scattered throughout its $\sim 1-$Mpc core region and well beyond.  
The globular cluster systems (GCSs) around these galaxies 
are well within reach
of the HST cameras, and thus give us an extraordinarily rich range
of target galaxies for comparative GCS studies.
Coma also has important implications 
for the determination of the Hubble constant
$H_0$, since its large recession velocity
($\sim 7100$ km s$^{-1}$) greatly exceeds any anticipated local peculiar 
motions, and it is situated at high galactic
latitude ($b = +87\fdg7$) nearly unaffected by foreground absorption,

{\ic} is a giant E2 galaxy located 
$\simeq 14'$ east in projection from the center of Coma
and has no luminous neighbors.
With an integrated magnitude 
$V_T^0 = 13.20$, it is the fifth brightest elliptical
galaxy within the $1\arcdeg$ central region of Coma. 
Thus, {\ic} presents a good opportunity to study the GCS of a
more-or-less ``normal'' large elliptical.
\cite{baum97} obtained the first 
deep WFPC2 exposures of \ic\ with exactly this purpose in mind, 
but they limited
their published analysis to only the photometry from the
PC1 frame on which the galaxy was centered.  However, considerably more
information on the radial structure of the GCS, its specific
frequency, and radial metallicity gradient are potentially 
available from the outer sections of the WFPC2 field.
In this paper, we re-analyze the archival HST/WFPC2 frames from
the \cite{baum97} program and discuss the global properties of
the \ic\ GCS using the entire body of data available in the
raw exposures.  

\section{DATA ANALYSIS}

The raw database comprises {\hst}/WFPC2 
observations taken on 1995 July 20/21 (Baum {\etal} 1997).
Eight images totalling 20500 s (essentially, a sequence of
full-orbit exposures) were taken through the $F606W$ (wide
$V$) filter, and two
images totalling 5200 s were taken through the $F814W$ ($I$) filter.
The techniques in our data analysis are similar to those
in Papers I and II of this series (\cite{kav99}; \cite{hk99}), in which we
analyzed the GCS around NGC 4874, the central cD giant.
The first step was to register
and combine the images to produce a master $V$ and master $I$
frame free of cosmic rays.  An excellent color composite image 
constructed from the combined frames is published by Baum et al.
For the PC1 frame, we produced a ``flattened'' 
image in which the overall elliptical contours of the central
galaxy were modelled and then subtracted.  In the WF2,3,4 fields,
we generated an empirical model of the galaxy light by median
filtering each frame and subtracting the smoothed image from the
original picture (after a preliminary star-finding and removal).

Our photometry was then carried out on the flattened master 
frames with the DAOPHOT II and ALLSTAR codes 
(\cite{ste92}). 
The instrumental magnitudes returned by ALLSTAR were transformed to the
standard $V$,$I$ system via the equations derived by \cite{holtzman95},
\begin{equation}
V = m(F606W) - {\overline {\Delta m}} + 22.093 + 0.254(V - I) + 0.012(V - I)^2 + 2.5\log(GR_i),
\end{equation}
\begin{equation}
I = m(F814W) - {\overline {\Delta m}} + 20.839 - 0.062(V - I) + 0.025(V - I)^2 + 2.5\log(GR_i),
\end{equation}
where $m(F606W,F814W)$ are the instrumental magnitudes returned
by ALLSTAR, and the $\overline {\Delta m}$ are constants
which shifts these ALLSTAR psf
magnitudes to the equivalent magnitude of the light within a 
$0\farcs5$ aperture.  The mean value of $\Delta m$ 
was determined empirically for each of the four WFPC2 fields
from roughly ten moderately bright, isolated stars in each CCD,
and applied to all our detected starlike images.  The typical internal
uncertainties in the mean $\Delta m$ values were $\pm 0.02$ mag.

It should be noted that at the distance
of Coma, all globular clusters are easily starlike in appearance
and thus these transformation
equations normalized to the $0\farcs5$ aperture 
can strictly be applied to them.
The calibration equation for $V$ quoted above is the synthetic model
transformation from Table 10 of \cite{holtzman95}, while the equation
for $I$ is the observationally based one from their Table 7.  If, instead,
we had chosen their Table 10 model transformation for $I$, our
resulting $(V-I)$ color indices would have ended up bluer by roughly
0.02 mag at the average color $(V-I) \simeq 1.2$ of the \ic\ globular
clusters (see below; this shift would have the effect of reducing the
deduced cluster metallicities by 
$\Delta$[Fe/H] = 0.1).\footnote{\cite{baum97} adopted the synthetic
transformation for $I$ from Table 10 of Holtzman \etal, but used the
specific curve for the color range $(V-I) < 1.0$.  
However, the majority of the
globular clusters are somewhat redder than this, so that the transformation
for $(V-I) > 1.0$ would have been preferable.  The net result is to
increase the small $(V-I)$ difference between their scale and ours 
by a further $\sim 0.02 - 0.03$ mag, although
our {\it instrumental} color scale $(F606W-F814W)$ agrees quite closely
with theirs.}   When we add the internal uncertainties in
the aperture corrections $\Delta m$, we
estimate that the zeropoints of either the magnitude or color scales
are uncertain to at least $\pm 0.03$ mag.

Finally, as an internal check on the relative zeropoints of the $V$ and $I$
scales between the PC1 chip and the three WF chips, we inspected the
mean $(V-I)$ color indices of the measured globular clusters in the
annulus around the center of IC 4051 ($R \sim
10'' - 20''$) that overlapped the PC/WF boundary.  We enforced the WF2,3,4
data in this annulus to have the same mean color as those in the PC1
region by adjusting the $I$ magnitudes (which are from 
much shorter exposures than $V$, thus 
internally more uncertain).  The final result places our $(V-I)$ scale
in agreement with the \cite{baum97} scale to within 0.03 mag, 
a level entirely consistent with the combined photometric uncertainties.

\begin{figure}
\epsscale{1.0}
\plotone{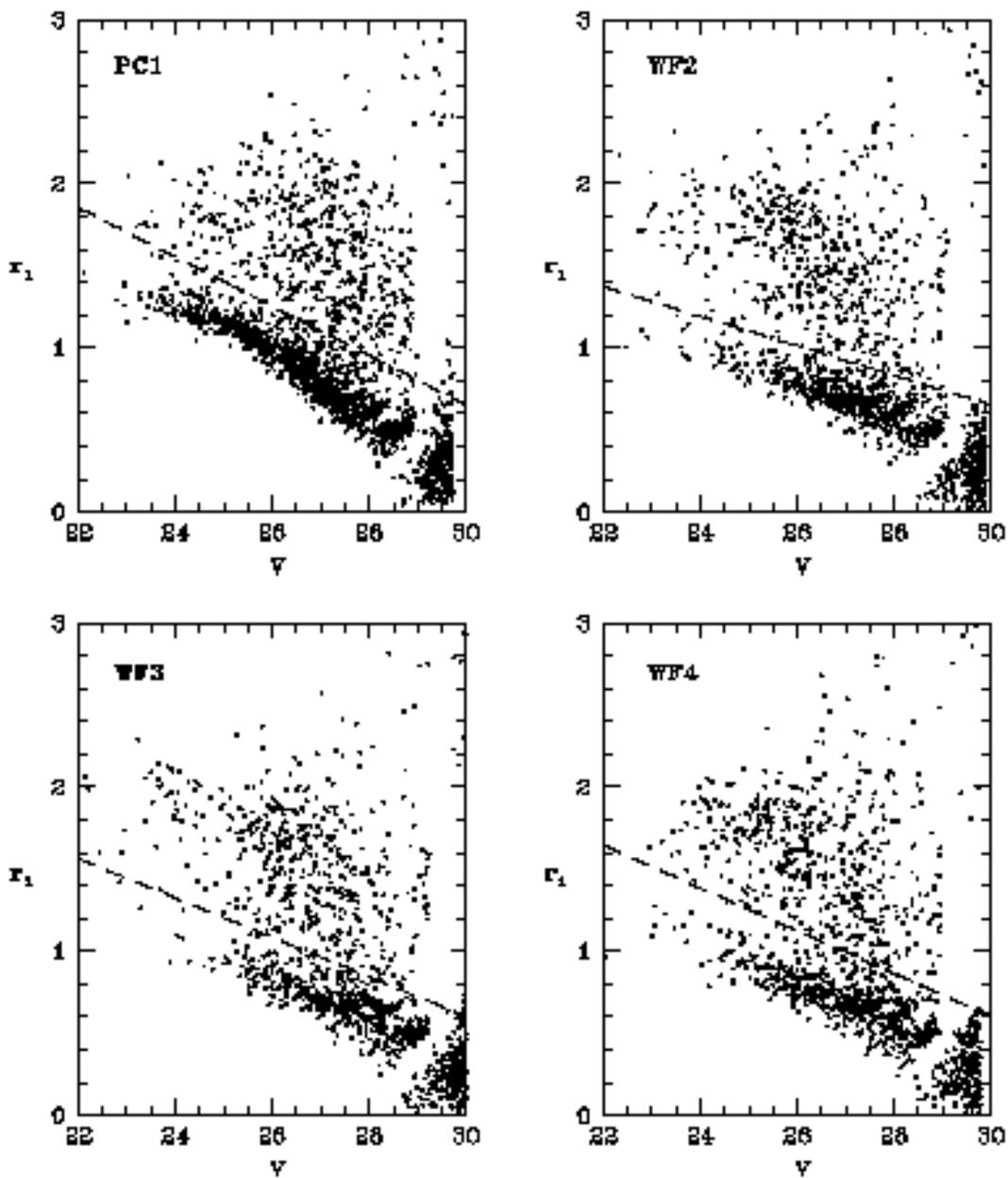}
\caption{Image classification plots for all
detected objects on the deep $V$ exposures.  The radial image moment
$r_1$ (see text) is plotted against $V$ for each of the four WFPC2
CCDs.  The dashed lines show the adopted boundaries separating
star-like objects (below the line) from nonstellar ones (above).  
The vast majority of the starlike objects are globular 
clusters around IC 4051.}
\label{fig1}
\end{figure}

As was already evident from the \cite{baum97} study, \ic\ has a
populous GCS, which appears as a very obvious swarm of faint objects
across the whole WFPC2 field.  The main source of sample
contamination is from very faint, compact background galaxies, with
a (nearly negligible) contribution from Galactic foreground stars.
A high proportion of the background galaxies can be eliminated
through conventional radial-moment image analysis.
For this purpose we used the $r_1$ radial moment as implemented
in \cite{hapv91}, 
\noindent
\begin{equation}
r_{1} = \left(\frac{\sum rI}{\sum I}\right),
\end{equation}
which is an intensity-weighted mean radius for the object calculated
over all pixels brighter than the detection threshold (see Harris
\etal\ 1991).  A straightforward plot of $r_1$ against magnitude
then shows a well defined stellar sequence, with nonstellar objects
scattering to larger $r_1$.  These classification
graphs for the four CCDs in $V$ are shown in
Figure \ref{fig1}.  The dashed lines indicate the adopted
cutoffs applied to the measurements.  Similar object
classifications were applied to the $I$ data (which, however,
have shallower limits), with 
the final culled data lists containing
4058 objects in $V$ and 1672 objects in $(V-I)$.

\begin{figure}
\epsscale{1.0}
\plotone{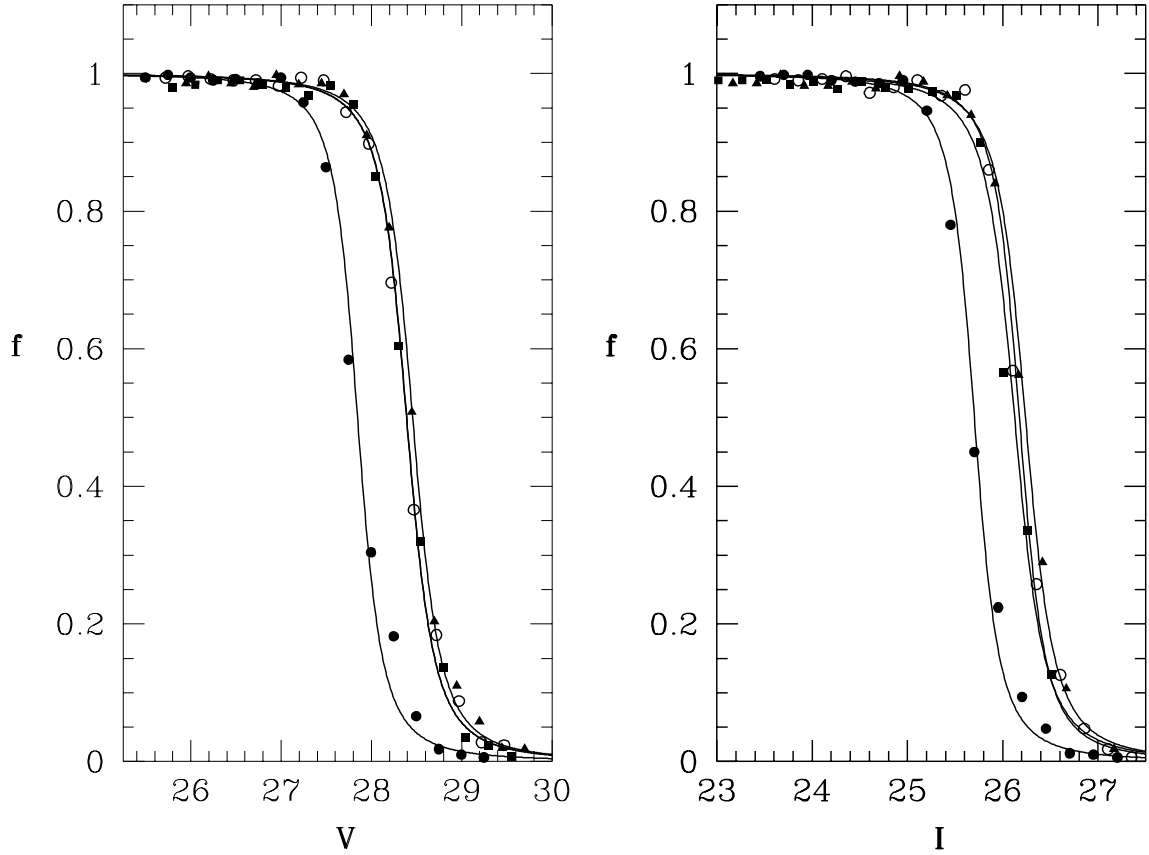}
\caption{
Completeness functions for the $V$ (F606W) and $I$ (F814W) photometry.
The solid dots (leftmost curve) represent the PC1 data, 
solid squares represent WF2, solid triangles WF3, and open circles WF4.
The lines through each set of points show the Pritchet interpolation
function curves described in the text and parametrized in Table 1.
The PC1 photometry has a noticeably brighter limiting magnitude.}
\label{fig2}
\end{figure}

The faint-end completeness of our photometry was
investigated through an extensive series of artificial-star 
tests on the master images.  The procedure performed here
was to add 500 artificial stars 
to each frame over a range of input magnitudes, measure these
frames through the normal DAOPHOT sequence, and find out how
many were recovered.  Fifteen of these trials were carried out,
with average resulting completeness fractions as 
displayed in Figure \ref{fig2}.
Convenient fits to the raw points are provided by the Pritchet
interpolation function,
\begin{equation}
f = \frac{1}{2} \left[{1 - 
	 \frac{{\alpha}(R-R_{lim})}{\sqrt{1+{\alpha}^2(R-R_{lim})^2}}}\right].
\label{compeq}
\end{equation}
Table \ref{tab1} summarizes the best-fit parameters to Equation 
\ref{compeq} for each CCD and bandpass.  In the Table,
($V_{lim}, I_{lim}$) are the magnitudes at which $f$ drops to 0.5, and the 
parameter $\alpha$ controls the steepness of the falloff.  
The curves for all three of the outer chips (WF2,3,4) are nearly
identical; for the inner PC1 chip, the limiting magnitudes are brighter,
driven by the spread of the PSF over many more pixels and (for $R \ltsim
10''$) the brighter background light.

\begin{figure}
\epsscale{1.0}
\plotone{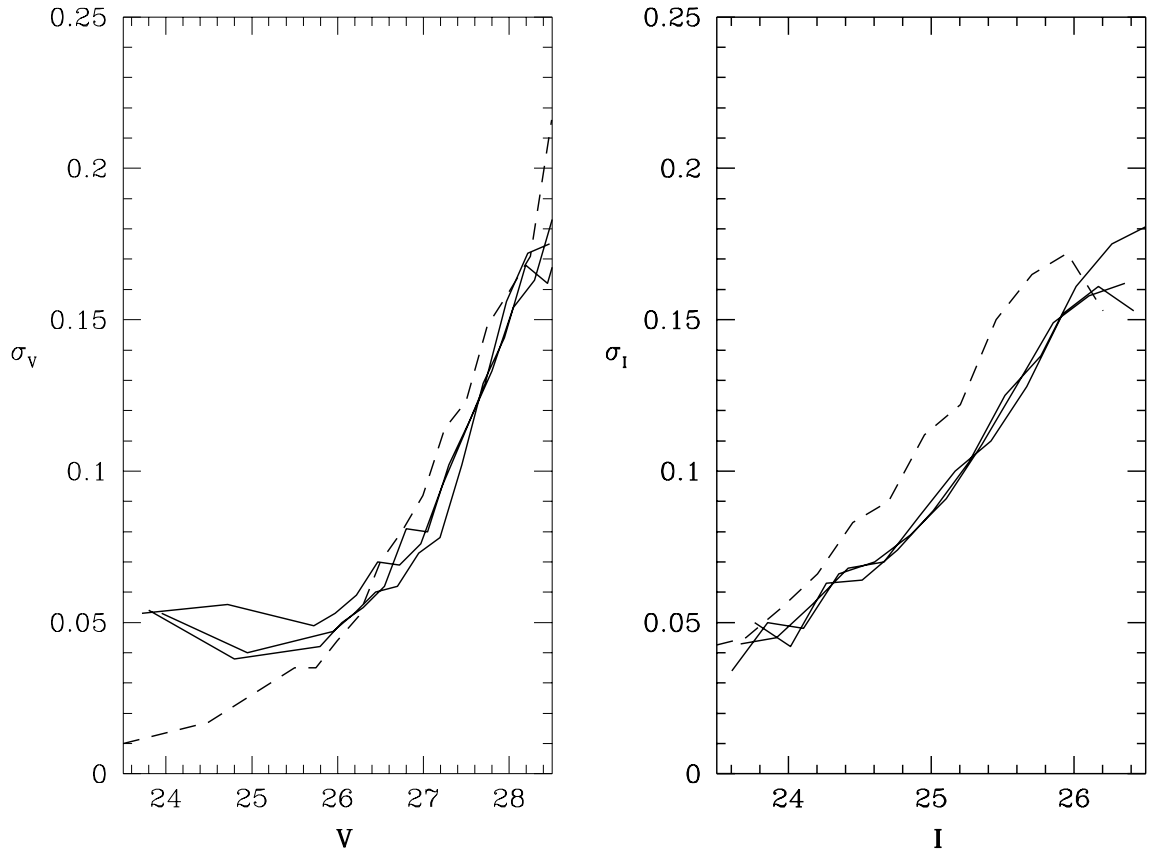}
\caption{
Mean photometric uncertainty as a function
of magnitude, derived from the artificial-star tests.  The dashed
line represents the PC1, while the solid lines are for the WF2,3,4
chips.  Here and in the completeness functions, there are no
significant differences among the three WF chips.}
\label{fig3}
\end{figure}

Lastly, Figure \ref{fig3} shows how the photometric measurement
uncertainties (also derived from the ADDSTAR completeness tests) increase
with magnitude.  At the formal limiting ($f=0.5$) magnitude, 
the rms uncertainty in the photometry reaches 0.15 mag.
Wherever possible, we avoid dealing with any features of the data
below that limit.

\section{COLOR AND METALLICITY DISTRIBUTIONS}

The distribution in colour of the globular clusters 
can be used to gain insight into the existence of
multiple sub-populations in the GCS.  Bimodal color distributions
are found about half the time in gE galaxies (e.g., \cite{kw99};
\cite{neil99}) and are often interpreted as relics of
at least two major phases of star formation in the early
history of the galaxy, whether by merger, accretion, or 
{\it in situ} processes.  With the conventional
``null hypothesis'' for giant ellipticals that the clusters 
are all old ($\gtsim 10$ Gy), the color index is primarily a tracer
of cluster metallicity.  
The colour-magnitude distribution for the 1672
objects measured in both $V$ and $I$ is shown in Figure \ref{fig4}.
At projected galactocentric radii larger than about $80 \arcsec$,
we found (see below) that the residual numbers of clusters dropped
nearly to zero, so we adopt this outer region as defining a
suitable ``background'' population.

To eliminate a few more contaminating objects, we reject
objects bluer than $(V-I) = 0.74$ or redder than
$(V-I) = 1.46$ (vertical dashed lines in Fig.~\ref{fig4}).
These colour limits generously include the range 
in colours of the 
known globular clusters in large galaxies (e.g., \cite{har96};
\cite{whi95}; \cite{neil99}).  We also further limited the color sample
to objects brighter than $V=26.0$ to ensure high completeness
at all colors.  

A ``clean'' color distribution for the GCS was then obtained 
by subtraction
of the background ($R > 80''$) color distribution, normalized
to the same total area as the inner population.
This procedure left a final total of 479 objects within the magnitude
and color limits given above, with a net distribution over $(V-I)$
as shown in the histogram of Figure \ref{fig5}.
The mean color of the sample is 
$\langle V - I \rangle = 1.12 \pm 0.01$ (internal uncertainty of the
mean), with a dispersion
of $\sigma_{V-I} = 0.13$.  
Subtracting an adopted foreground reddening $E(V-I) = 0.014$ and
using the calibration of $(V-I)_0$ in terms of metallicity
given in Paper II, 
$$ (V-I)_0 = 0.17 \, {\rm [Fe/H]} \, + \, 1.15 \, , $$
we then estimate that the IC 4051 GCS as a whole has 
$\langle$Fe/H$\rangle \simeq -0.3$.  The peak position of this
color distribution is quite similar to the metal-{\it rich} components
in other giant ellipticals such as NGC 4472 (\cite{geisler96}), M87
(\cite{whi95}; \cite{kun99}), and other Virgo members (\cite{neil99}).
However, the metal-{\it poor} component which is usually found in
these same galaxies at a mean color $(V-I) \sim 0.95$ 
or [Fe/H] $\sim -1.5$ (and which we
found in the Coma cD NGC 4874; see Paper II) is entirely missing
in IC 4051, or at very most is a fringe component buried in the
wings of the main distribution.

We cannot place firm
limits on the intrinsic dispersion $\sigma$[Fe/H], since the mean
observational measurement scatter over the sample is $\sigma_{V-I} 
\simeq 0.11$, comparable with the observed sample dispersion of $\pm0.13$.
Nevertheless, subtracting off the observational scatter in quadrature,
we estimate roughly $\sigma_0$[Fe/H] $\simeq 0.4$, which
is in close agreement (for example) with the value $\sigma$[Fe/H] = 0.38
found by Geisler \etal\ (1996) \nocite{geisler96} 
to fit each of the metal-rich and
metal-poor components in NGC 4472.  In the Milky Way, the well known
bimodal MDF has been found to be fit by Gaussian functions with
dispersions near $ 0.3$ dex (\cite{zin85}; \cite{armandroff88};
\cite{har99}).
For IC 4051, a single Gaussian with the same mean and
standard deviation as the sample (Fig.~\ref{fig5}) matches
the MDF with a $\chi^2 \simeq 14.6$ over 14 degrees of freedom, 
which provides no strong evidence for bimodality (but see below).

Trends of mean color with either galactocentric distance or
magnitude were also searched for.
Table \ref{tab2} shows the mean color and dispersion in 0.5-magnitude bins
from $V = 22.5$ to 26.0.  These binned means reveal
no significant change in color with luminosity.
However, slightly more interesting features emerge in the graph
of color versus radius (Figure \ref{fig6}).
Binned mean colors, listed in Table \ref{tab3}, indicate no 
systematic change in color for $R \gtsim 10''$, but within $10''$
the clusters are indeed slightly redder
than the overall mean.  The distribution
in its entirety is barely suggestive of two sub-populations:  one
centered on $(V-I) \sim 1.2$ which is found at all radii; and a second,
slightly bluer one centered near $(V-I) \sim 1.0$.  The lack of bluer
clusters within $R \ltsim 10''$ is then largely responsible for the
inner color gradient of the whole sample mentioned above.
Much stronger versions of this same effect have shown up in 
some other giant E or cD galaxies
with far more obvious bimodal MDFs
(e.g., \cite{secker95};
\cite{geisler96}; \cite{lee98}; \cite{ostrov98}).
In these, the different central concentrations of the metal-rich
and metal-poor subsystems produce a steady outward change in
the relative proportions of blue-to-red clusters with radius
and thus a mean metallicity gradient.

Using Fig.~\ref{fig6} as a guide, we divided the sample of objects
at $(V-I) = 1.07$ and tested the radial distributions of the
bluer and redder halves.  A standard Kolmogorov-Smirnov two-sample
test indicated that their spatial distributions are significantly
different (the redder half is more centrally concentrated) at
the 99\% level, suggesting to us that the inner gradient is indeed
a real effect.

We therefore {\it very tentatively suggest} that the IC 4051 system
may contain a bimodal MDF in which the two modes are rather closely
spaced in mean metallicity, thus heavily blurred out by the raw
photometric measurement uncertainty.  Numerical experiments with various
two-component fittings of the entire MDF lead to models of the form shown
in Figure \ref{fig7}.  Here, a sample twin-Gaussian fit is shown in
which the bluer (metal-poor) component is centered at $(V-I)=1.00$
or [Fe/H] $\simeq -0.96$,
the redder (metal-rich) one at $(V-I)=1.17$ or [Fe/H] $\simeq +0.04$, 
both have dispersions
$\sigma(V-I) = 0.10$, and the redder one contains about 55\%
of the total sample.  The combined components now represent the 
total shape of the MDF better, with its modest skewness toward the red
side (the total $\chi^2$ is 12.6).  The relative proportions of blue
and red components, however, are quite uncertain (the formal
uncertainties are $\pm0.1$, but variations of factors
of two in the proportions give scarcely different overall fits).

Clearly, this particular two-component 
model is only illustrative of the range of possibilities:
the moderately small difference in color between the two components, and the
very significant broadening of the MDF by photometric measurement
uncertainty, do not justify more extensive analysis.
However, it would clearly be of value to
measure the MDF of this populous globular cluster system with a
photometric index much more sensitive to metallicity than $(V-I)$,
in which the subpopulations would be far more clearly revealed.
A more sensitive color index would also permit establishment of the
true mean [Fe/H] with much less zeropoint uncertainty.

Lastly, it is worth comparing the mean colors of
the GCS components to that of the halo {\it light} of
the central galaxy.  \cite{meh98} find $(V-I) \simeq 1.30$
at a projected radius $R = 10''$, increasing inward to $(V-I) = 1.35$
at the very center.  This color range is distinctly redder than
the typical levels $(V-I) \simeq 1.20 \pm 0.03$ 
for giant E galaxies (\cite{but95}; \cite{pru98}).  
The measured absorption line indices (Mg, Fe, H$\beta$) 
lead \cite{meh98} to conclude, in line with the integrated
color, that the core of IC 4051 is extremely old and very metal-rich,
perhaps as high as [Fe/H] $=+0.25$.  However, the 
deduced metallicity from the line indices becomes
lower at larger radii, dropping to an equivalent [Fe/H] $\sim -0.5$
for $R \gtsim 20''$ (the effective radius $r_e$ of the light profile),
similar to the inner GCS.
Mehlert \etal\ find that IC 4051 harbors
an old, co-rotating core with an unusually large ``break radius'' (it is
detectable out to $5''$ or 3.4 kpc) but which contributes $\ltsim$ 1\%
of the total light of the galaxy.  If this inner stellar disk
is a signature of a dissipational merging event, it is likely to have
occurred at early times.

\section{THE LUMINOSITY DISTRIBUTION}

As \cite{baum97} showed, the $V$ photometry reaches faint
enough to reveal the ``turnover point'' (peak frequency)
in the globular cluster luminosity function (GCLF).
By adding in the photometry from the WF chips, we have been
able to double the total sample of clusters and thus improve
the definition of the GCLF.

The distribution of all the detected objects classified
as ``starlike'' and used to define the GCLF is shown in Figure \ref{fig8}.
These are, quite evidently, strongly concentrated to the
center of IC 4051 (much more so than in the GCS of
the Coma supergiant NGC 4874; see Papers I and II).  
More or less arbitrarily, we take the region $R > 80''$
marked by the outer dashed line in Fig.~\ref{fig8},
as defining the luminosity function of the background population,
to be subtracted statistically from the inner ($10'' < R < 80''$)
zone after correction for photometric incompleteness.

The results of this exercise for each of the four CCD chips
separately are shown in Figure \ref{fig9}.  Aside from the
noticeably brighter completeness limit for the PC1 zone,
no significant differences in the GCLF shape or turnover from
place to place are evident.  (The GCLF peak for the PC1 region
shows an apparent peak fainter than $V \sim 28$, but this
is fainter than
the 50\% completeness limit and so cannot be given much weight.)
We therefore add all four sectors
to form the composite GCLF shown in Figure \ref{fig10}.
The numerical results in 0.3-mag bins are listed in Table \ref{tab4}:
here, successive columns give (1) the $V$ magnitude range of
the bin (2) the number of detected starlike objects in the
inner ($10''-80''$) zone (3) the number in the outer ($> 80''$)
background zone (4) the number in the inner zone corrected for
completeness, and (5) the net GCLF, after subtraction of the
area-normalized background counts.

To estimate the turnover level and shape of the GCLF, we fit 
a standard Gaussian interpolation function (\cite{har91}; \cite{jac92})
to the data shown in Fig.~\ref{fig10}, setting the 
standard deviation $\sigma_V$ of the curve and then solving for
the best-fit turnover level $V^0$.  Trials with different
adopted $\sigma_V$'s gave the results summarized in Table \ref{tab5}.
The reduced $\widetilde{\chi}^2$ values favor a solution in the
broad range $\sigma_V \sim 1.4 - 1.8$, with little to choose
among values in this range 
in a formal sense.  However, it is well known that both
$\sigma_V$ and $V^0$ tend to be overestimated in situations like
these where the magnitude limit of the data reaches barely past the
actual turnover (e.g., \cite{han87}; Paper I) since the solutions
for the two parameters are correlated.  For this reason, we favor
a choice in the narrower range $\sigma_V \simeq 1.4 - 1.6$ and
$V^0 \simeq 27.6 - 28.0$.  Sample Gaussian curves for the extremes
of this range are shown in Fig.~\ref{fig10}.

Our final adopted pair of parameters is $V^0 = 27.8 \pm 0.2$, 
$\sigma_V = 1.5 \pm 0.1$.  For comparison, \cite{baum97} found
$V^0 = 27.72$ employing a different and more complex fitting function.
As is discussed more extensively
in Paper I, this turnover level is also similar to what we found in
the central cD NGC 4874.  Using both of them combined, along with
a calibration of the absolute magnitude of the turnover point based
on the Virgo ellipticals, we find $d \sim 100$ Mpc for Coma along
with a Hubble constant $H_0 \simeq 70$ (see Paper I).

A second and more physically oriented way to display the same material
is as the {\it luminosity distribution function} (LDF), or number
of clusters per unit (linear) luminosity.  (The relation between
the GCLF and LDF forms is exhaustively discussed by \cite{mcl94}.)
The LDF is shown in 
Figure \ref{fig11}.  At levels brighter than the GCLF ``turnover''
(which in turn is only slightly brighter than the photometric
completeness limit), the LDF clearly approximates a power-law
falloff toward higher luminosity, $N(L) dL \sim L^{-\alpha}$.
To second order, however, the slope $\alpha = -d$log($N$)/$d$log($L$)
appears to steepen slightly at the upper end:
an unweighted least-squares fit to all bins brighter than the
turnover yields $\alpha = 2.05$, while exclusion of the half-dozen
very brightest bins yields $\alpha = 1.75$.

These power-law forms -- as well as logarithmic slope values $\alpha \sim 2$ --
are entirely similar to what has been found in a wide range of
other galaxies from dwarf ellipticals to spirals (\cite{hp94};
\cite{dur96}).  However, in most giant ellipticals studied to date, the
slopes tend to be somewhat flatter at $\alpha \sim 1.5 \pm 0.3$ (\cite{hp94}).
The total shape for log $(L/L_{\odot}) \gtsim 5$, complete with
its progressive steepening toward higher luminosity, 
can be well matched by a formation model
in which protocluster clouds build up by collisional agglomeration
and in which the more massive clouds have shorter lifetimes against
star formation (\cite{mp96}; \cite{har99}).  
Our data for IC 4051 add further to the general body of material
which indicates a remarkable place-to-place similarity in the
luminosities of old globular clusters, and thus a quasi-universal
formation process.

\section{RADIAL DISTRIBUTION AND SPECIFIC FREQUENCY}

Because the GCS around IC 4051 is quite centrally concentrated
(see Fig.~\ref{fig8}), we can use the complete WFPC2 data to
define the spatial distribution outward nearly to its limits.
The radial profile of the raw counts for all starlike objects
brighter than $V = 27.0$, for which the data are highly complete
nearly in to the central core of the galaxy, is shown in Figure \ref{fig12}.
The inner core ($R \ltsim 5''$), in which the projected density of
clusters is nearly flat, continues outward to a steep power-law
falloff which covers most of our survey area.  Finally,
for $R > 80''$, the number density $\sigma$ begins to level off
towards its eventual far-field background level; more or less
arbitrarily, we set this background at $\sigma_b = (0.02 \pm 0.01)$
arcsec$^{-2}$ as representing
nearly the average of the outermost two points.
(As will be seen below, small
differences in the adopted $\sigma_b$ 
will not have major effects on any of our subsequent conclusions.)

The complete profile data broken into circular annuli are listed
in Table \ref{tab6}, giving the number of objects in each bin, the surface
area of the annulus, and the projected density $\sigma$.
The residual number density of clusters, $\sigma_{cl} = \sigma-\sigma_b$, 
is plotted in Figure \ref{fig13}.  Simple \cite{king66} models can
be fitted to the $\sigma_{cl}$ data points to give rough estimates
of the GCS core radius and central concentration:  performing a weighted
fit in  the manner described in Paper II, and ignoring the very uncertain
outermost three points, we find a formal best-fit
core radius $R_c = 10\farcs25$ (equivalent to 5.1 kpc at the adopted
Coma distance) as well as a concentration index $c = 1.45$ for a dimensionless
central potential $W_0 = 6.26$.  The core radius is four times
smaller in IC 4051 than the $\sim 22$ kpc
value we found in the much more extended NGC 4874 (Paper II).

A second comparison can be made with the halo {\it light} of the galaxy.
It is conventionally found in giant ellipticals
that the GCS is a more spatially extended
system as a whole than the halo (\cite{har91}, 1999).  IC 4051 is no
exception, despite its overall compact structure.  
In Fig.~\ref{fig13}, we show the 
wide-field surface intensity profiles
in $\mu_R$ measured by \cite{str78} and \cite{jfk92}.  Although the
Strom data are photographically measured, their profile agrees tolerably with
the more recent CCD measurements of \cite{jfk92} over their region
of overlap.

The bulk of the GCS profile is more extended than the halo light,
except possibly for the outer ($R \gtsim 30''$) regions where their
slopes are more nearly similar.   For $R \gtsim 20''$,
the GCS profile behaves as $\sigma_{cl} \sim R^{-2}$, although
at the largest radii little weight can be placed on
the very uncertain outermost half-dozen points.  There is a strong
hint from the halo light profile that the galaxy may be truncated
past $R \sim 60''$ (about 30 kpc), though here again the profile
is very sensitive to slight differences in the adopted background level,
so not much meaning can be ascribed to the slope differences 
between the GCS and the halo there.
For giant E galaxies in general, a rough mean relation between
galaxy luminosity $M_V^T$ and the radial falloff outside the
central core is (\cite{kai96}) $d$log$\sigma$/$d$log$R$ $\simeq -0.29 M_V^T
- 8.00$.  For IC 4051, this relation would predict $\sigma_{cl} \sim
R^{-1.65}$, somewhat flatter than the observed $R^{-2}$ trend.

Calculating the total GCS population and specific frequency is now
a straightforward matter.  From Table \ref{tab6}, we multiply $\sigma_{cl}(R)$ 
by the area of each annulus, then sum the annuli to get the total cluster
population out to the limits of our survey.  We find
$N = (1845 \pm 165)$ for $V \leq 27.0$ and $R \ltsim 130''$.
If the true GCLF turnover magnitude is at $V^0 = 27.8 \pm 0.2$ (see above),
then we must multiply this raw total by $\simeq (3.35 \pm 0.52)$ to
estimate the total cluster population over all magnitudes, giving
$N_{cl} = 6180 \pm 1100$.  The integrated luminosity of the galaxy
is $V^T = 13.20$ (RC3 catalog value), corresponding to $M_V^T = -21.9$
for our adopted Coma distance.  Thus, the specific frequency is
$$ S_N = N_{cl} \cdot 10^{0.4(M_V^T + 15)} = 10.8 \pm 1.9 \, .$$

In strict terms this is a {\it lower limit} to the true global
$S_N$, since we have not accounted for any cluster population outside
the $\simeq 120''$ radial limit of our WFPC2 field.  However, given
that the halo is clearly declining quite steeply in this region
(Fig.~\ref{fig13}), any such population correction is likely to be
small.  A generous but reasonable 
{\it upper limit} estimate to the total population can
be made if we assume that the GCS profile continues as $\sigma_{cl}
\sim R^{-2}$ outward to the nominal tidal radius at $R_t \sim 230''$.
This assumption gives an additional $\sim 380 \pm 300$ clusters brighter than
$V = 27$, which then translates to $S_N = 12.6 \pm 2.6$.

Placing more weight on the lower limit -- which reflects 
the steep falloff of the system near the radial limit of
our data -- we adopt a final estimate
$$S_N{\rm (final)} = 11 \pm 2 \, .$$

Remarkably, this GCS population ratio is
several times higher than the $S_N \ltsim 2$ value found in
NGC 4881 (\cite{baum95}), a galaxy which is quite comparable with
IC 4051 in luminosity, structure, and location on the outskirts
of the Coma core.  This high $S_N$, in fact, 
places IC 4051 in the range which
is conventionally reserved for the central-giant cD galaxies
like M87 and many other BCGs (\cite{har98}; \cite{bla97}, \nocite{bla99}
1999).
It is, perhaps, particularly noteworthy that IC 4051 has a specific
frequency three times higher than the central cD {\it in its
own host galaxy cluster}, NGC 4874 (see Paper II).  No other instance
of such a large contrast between a low$-S_N$ central cD and 
a higher$-S_N$ outlying
elliptical is known.  IC 4051 provides striking evidence that a central
location in a rich cluster environment is {\it not} required to form
a high population of globular clusters.

\section{DISCUSSION}

A brief summary of our findings for IC 4051 is that its GCS is
(a) almost entirely metal-rich, albeit possibly with two narrowly
separated subcomponents;
(b) relatively compact in radial structure; and (c) a ``high specific
frequency'' system despite that fact that its host galaxy is not a
central giant elliptical nor one with a cD-type envelope.

Just as in Paper II for NGC 4874, we now attempt to use the integrated
characteristics to reconstruct a partial history of the system.
Formation scenarios for giant ellipticals tend to fall into three
basic camps:  (a) ``in situ'' formation, whereby the galaxy
condenses by dissipative collapse of gas clouds in its immediate
vicinity, in one or more major bursts; (b) later mergers of pre-existing
disk-type galaxies with both gas and stars; or (c) successive mergers
or accretions of smaller gas-poor satellites.  Various combinations of
these extremes are, of course, possible, and even likely.

For IC 4051, the lack of {\it low-}metallicity clusters already places
fairly strong constraints on the range of possible formation events.
For example, the mechanism investigated by \cite{cot98} -- in 
which an original metal-rich ``seed'' gE accretes dozens or hundreds of
smaller satellites -- is unlikely, since these dwarf satellites would
have brought in a population of hundreds or even thousands of 
low-metallicity clusters, which we do not see.  

Similarly, merger-formation 
models in which gas-rich disk galaxies combine to build a composite
elliptical (\cite{ash92}) would predict a strong component of metal-poor 
clusters in the resulting MDF from the globular clusters that were
present in the pre-merger galaxies.  These merger models also have
severe difficulty in generating high specific frequency products,
since increasing the cluster population {\it relative to the field stars}
by a large enough amount to produce high $S_N$ requires 
very large ($> 10^{10} M_{\odot}$)
input gas masses, more than is routinely available in disk galaxies today.
The normal merger route does appear to be quite effective as a logical
source for low$-S_N$ field ellipticals (see \cite{har99} or \cite{whi95a}
for much more extensive discussion).  

However, if either the merger or accretion 
processes are taken to an extreme form
in which {\it the merging objects are almost completely gaseous}, then they
become closely similar to the {\it in situ} route, and the conundrum
of the missing low-metallicity clusters can be more easily circumvented.
If the gas supply -- however it was assembled -- underwent most or all
of its star formation in the high-pressure, high-density environment
of the protoelliptical, then the conversion of gas to stars
would have run much further to completion
and built up the metallicity to the high levels that we now observe.
Later gaseous mergers are, of course, not ruled out:  the central
corotating disk in the core of IC 4051 (Mehlert \etal\ 1998)
\nocite{meh98} with its very high metallicity is a likely signature
of such an event, though at its $\ltsim 1$\% contribution to the
present-day luminosity, it probably did not form more than a few dozen
globular clusters along with it, and even these would have mostly
disrupted by now if they resided in the central few kpc of the core.

The relatively compact structure of the galaxy may be the result of
tidal trimming (``harrassment'') from the Coma potential well (e.g.,
\cite{moo96}).
The radial velocity of IC 4051 (4940 km s$^{-1}$) is almost two standard
deviations away from the
Coma centroid (6850 km s$^{-1}$; see \cite{col96}), 
indicating that this galaxy
oscillates back and forth through the cluster and   
is now passing through the dense Coma core at high speed.

These elements of an evolutionary scenario for IC 4051 are in strong
contrast to NGC 4874, for which we argued (Paper II) that 
a large fraction of its clusters (which are
almost entirely low-metallicity) could
have been acquired by accretions of smaller satellites.
In IC 4051, we are forced to argue that the bulk of its clusters
formed {\it in situ}.   The globular clusters in these two galaxies
provide unique evidence for the view that large E galaxies can form by
radically different evolutionary routes.

One of the most challenging elements of IC 4051 to interpret 
is certainly the high specific frequency of its GCS.
In the previous literature (\cite{har91}, 2000; \cite{bla97}, 1999;
\cite{har98}; \cite{mcl99}) it has become conventional to associate
high $S_N$ with giant galaxies at the centers of rich clusters.
These central BCG's or cD's can have had histories
of star and cluster formation through inflowing gas clouds and filaments,
mergers, and accretions (e.g., \cite{dub98}) that were much more
extended than for normal outlying ellipticals.  Recently, the view
has been developed that such high$-S_N$ galaxies should be regarded
not as ``cluster-rich'' but rather as ``star-poor''
(\cite{bla97}, 1999; \cite{har98}; \cite{mcl99}).  In this scheme,
we postulate that 
the protogalactic gas started forming globular clusters at early times
at a normal efficiency rate,
but was then disrupted (perhaps by supernova-driven 
galactic winds, or by tidal shredding during infall;
cf.~the papers cited above) before its star formation
could run to completion.  The leftover gas now remains around these
galaxies as their hot X-ray halos.  This picture, however, assumes  
that the globular clusters form earlier than the bulk of the field stars
in any given round of star formation -- not an implausible requirement
given the bulk of the observational evidence for starburst systems
(see Harris 2000) and given that globular clusters emerge from the densest,
most massive protocluster clouds.

McLaughlin (1999) \nocite{mcl99} defines a globular
cluster formation efficiency, measured empirically as the mass ratio
$$ \epsilon = {M_{cl} \over {M_{\star} + M_{gas}} } $$
where $M_{\star}$ and $M_{gas}$ are the masses 
within the galaxy in the form of visible stars and
in the X-ray gas respectively.  He finds that $\epsilon$ is essentially
identical in the well studied Virgo and Fornax
systems M87, NGC 4472, and NGC 1399 (despite their very different
$S_N$), providing evidence for a
``universal'' globular cluster formation efficiency $\epsilon \simeq
0.26$\% relative to the
{\it initial protogalactic gas supply}.  The total mass ratio
$\epsilon$ is a more important indicator of cluster formation
than $S_N$, which is only a measure
of the cluster numbers (or equivalently total mass) relative to 
the galaxy light.  In other words, $S_N$ is a measure of
only the gas mass $M_{\star}$ that got converted to stars.
Additional support for the near-universality of 
$\epsilon$ in several other BCG's is found by Blakeslee (1999) \nocite{bla99}.

In this view, {\it any high$-S_N$ galaxy
should then be surrounded by a massive X-ray gaseous component}
whether or not it is a centrally dominant giant. 
Notably, IC 4051 is indeed one of the
few Coma ellipticals with an individually detected X-ray halo.
Dow \& White (1995), \nocite{dow95} 
from ROSAT observations of the Coma core region, find
that IC 4051 is detectable at the $2-\sigma$ level in the soft
X-ray range $0.2 - 0.4$ keV, but not in the higher $0.4 - 2.4$ keV range.  
If it were at the $\sim 6.3$ times closer distance of Virgo, IC 4051 
would have a total $L_X \simeq 5 \times 10^{41}$ erg s$^{-1}$.  This
level makes it quite comparable with the
Virgo giant NGC 4472, which has $L_X \simeq 6 \times 10^{41}$ erg s$^{-1}$
in the soft X-ray regime (\cite{fab92}; \cite{irw96}; \cite{mat97};
\cite{buo98}).  However, this amount of X-ray gas corresponds to only
$\sim 5$\% of the stellar mass (\cite{mcl99}) and NGC 4472, as expected,
has only a ``normal'' specific frequency level $S_N \simeq 5$.

With our adopted distance ratios for Virgo and Coma, we find that
IC 4051 is about half as luminous as NGC 4472, so if it has a roughly
similar amount of X-ray gas mass, this gas would only make up
$\sim 10$\% of its stellar mass.  Along with $S_N \simeq 11$,
we find that these parameters convert to
a {\it present-day} value for the mass ratio in IC 4051 of
$\epsilon \sim 0.005$, twice as large as McLaughlin's (1999)
\nocite{mcl99} fiducial value.

Nominally, it therefore seems that IC 4051 acts against the
paradigm of a universal globular cluster formation efficiency.
An obvious possibility, however, is that IC 4051 originally
did possess much more gas shortly after its main era of
globular cluster formation, but that most of this unused material was
quickly stripped away 
as IC 4051 went through its first few passages of the Coma core.
This gas would have joined the general reservoir 
of hot gas spread throughout the Coma potential well.
The same mechanism which resulted in this galaxy's compact structure
might then have plausibly left it with the unusual combination of 
high $S_N$ and modest amount of X-ray gas that we now see.

A situation which would act much more strongly to falsify 
McLaughlin's case for a
universal $\epsilon$ would be the opposite one:
that is, a galaxy with a massive X-ray halo but a ``normal''
or subnormal $S_N \ltsim 4$.  In  such a case it would be much
harder to avoid the conclusion that the formation efficiency
of globular clusters was genuinely different (and low).
Does the central Coma giant NGC 4874 present us with such a case?
As we found in Paper II, 
NGC 4874 is {\it not} a high-$S_N$ system and is embedded within
a very massive X-ray envelope.  This X-ray gas is, however, so extended
that must belong to the general Coma potential well as a whole, with no
detectable concentrated component that 
can be associated with NGC 4874 itself
(\cite{dow95}).  Thus there are ambiguities in
the interpretation that are hard to circumvent.  Better candidates
would be E galaxies with massive X-ray halos that are not at
the centers of rich clusters.

Finally, we may compare the interesting case of IC 4051 with that
of its Coma neighbor NGC 4881 (\cite{baum95}), a giant E galaxy
of similar location, size, and structure.  Curiously, NGC 4881 holds a
GCS of {\it low} specific frequency ($S_N \ltsim 2$) which appears
to be almost entirely metal-{\it poor}, just the opposite of
IC 4051.  It has no significant amounts of X-ray gas (\cite{dow95}).
We speculate that NGC 4881 may have resulted from the
merger of smaller galaxies in which these metal-poor globulars had
already formed.  These mergers should have been rather gas-poor to
prevent the formation of newer and more metal-rich clusters.
This is, however, an extremely sketchy interpretation, and there is an
obvious problem with the much higher metallicity of the host galaxy
light (how did the bulk of the giant E galaxy form at higher metallicity
without leaving behind some metal-rich globular clusters?  See Paper II
for additional discussion).  

The Coma ellipticals clearly present a wide range of GCS characteristics
that strongly challenge the array of current galaxy formation models.

\acknowledgments

This research was supported through a grant to WEH from
the Natural Sciences and Engineering Research Council of Canada.
We thank Peter Macdonald of Ichthus Data Systems and the Department of
Mathematics and Statistics at McMaster 
for providing the MIX multicomponent fitting code.
Bill Baum provided several constructive comments on the first version
of the text.

\clearpage


\begin{deluxetable}{ccccc}
\tablenum{1}
\tablecaption{Completeness Function Parameters  \label{tab1}}
\tablewidth{0pt}
\tablecolumns{5}
\tablehead{
\colhead{CCD} & \colhead{$F606W (V)$} & \colhead{ } &
\colhead{$F814W (I)$} & \colhead{ }  \nl
\colhead{ } & \colhead{$\alpha$}  & \colhead{$V_{lim}$}  & \colhead{$\alpha$}  & \colhead{$I_{lim}$ }
}
\startdata
PC1   &   3.508 &  27.85     & 3.682    & 25.70 \nl
WF2   &   3.141 &  28.39     & 3.078    & 26.13\nl
WF3   &   3.123 &  28.46     & 3.157    & 26.24\nl
WF4   &   3.154 &  28.39     & 3.530    & 26.18\nl
\enddata
\end{deluxetable}

\clearpage

\begin{deluxetable}{crcc}
\tablenum{2}
\tablecaption{Mean Color vs. $V$ Magnitude  \label{tab2}}
\tablewidth{0pt}
\tablecolumns{4}
\tablehead{
\colhead{$V$} & \colhead{N } & \colhead{$\langle V - I \rangle$} & 
\colhead{$\sigma_{(V-I)}$} 
}
\startdata

$      22.5  \!-\!    23.0  $ &   7 &    1.058  &   0.031 \nl
$      23.0  \!-\!    23.5  $ &   7 &    1.136  &   0.053 \nl
$      23.5  \!-\!    24.0  $ & 20  &   1.101   &  0.018  \nl
$      24.0  \!-\!    24.5  $ & 43  &   1.111   &  0.016  \nl
$      24.5  \!-\!    25.0  $ & 77  &   1.128   &  0.012  \nl
$      25.0  \!-\!    25.5  $ & 145 &    1.113  &   0.010  \nl
$      25.5  \!-\!    26.0  $ & 180 &    1.119  &   0.010  \nl

\enddata
\end{deluxetable}

\clearpage

\begin{deluxetable}{lrcccc}
\tablenum{3}
\tablecaption{Mean Color vs. Radius  \label{tab3}}
\tablewidth{0pt}
\tablecolumns{6}
\tablehead{
\colhead{R (arcsec) } & \colhead{ N } & \colhead{$\langle V - I \rangle$ } & 
\colhead{$\pm$ } & 
\colhead{$\langle {\rm Fe/H} \rangle$ } & \colhead{$\pm$ }
}
\startdata
$\phn0.0  \!-\!\phn 8.0$ & 79&1.170   &    0.013  &  $  +0.04 $ & 0.08 \nl
$\phn8.0  \!-\! 12.0$ & 80 &    1.141   &    0.011  & $ -0.14 $& 0.07 \nl
$12.0 \!-\! 16.0$  &  61   &    1.116   &    0.014  &  $  -0.28 $   &    0.08 \nl
$16.0 \!-\! 26.0$  &  68   &    1.083   &    0.015  &  $  -0.48 $   &    0.09 \nl
$26.0 \!-\! 40.0$  &  75   &    1.120   &    0.013  &  $  -0.26 $   &    0.08 \nl
$40.0 \!-\! 55.0$  &  51   &    1.120   &    0.017  &  $  -0.26 $   &    0.10 \nl
$55.0 \!-\! 80.0$  &  65   &    1.050   &    0.015  &  $  -0.67 $   &    0.09 \nl

\enddata
\end{deluxetable}

\clearpage

\begin{deluxetable}{cllll}
\tablenum{4}
\tablecaption{Binned GCLF Data  \label{tab4}}
\tablewidth{0pt}
\tablecolumns{4}
\tablehead{
\colhead{$V$} & \colhead{N$_{in}$}& \colhead{N$_{out}$ } & \colhead{N$_{in,corr}$} 
& \colhead{N$_{final}$}
}
\startdata
$ 22.1 \!-\! 22.4$ & \phn\phn  1.0 $\pm$  1.0 & \phn 0.0   $ \pm$  0.0  &  \phn\phn  1.0 $\pm$   1.0 & \phn\phn   1.0 $\pm$   1.0 \nl
$ 22.4 \!-\! 22.7$ & \phn\phn  1.0 $\pm$  1.0 & \phn 1.0   $ \pm$  1.0  &  \phn\phn  1.0 $\pm$   1.0 &\phn\phn \llap{$-$}0.9 $\pm$   1.5 \nl
$ 22.7 \!-\! 23.0$ & \phn\phn  5.0 $\pm$  2.2 & \phn 1.0   $ \pm$  1.0  & \phn\phn  5.0 $\pm$   2.2 & \phn\phn   3.1 $\pm$   2.5 \nl
$ 23.0 \!-\! 23.3$ & \phn\phn  1.0 $\pm$  1.0 & \phn 0.0   $ \pm$  0.0  & \phn\phn  1.0 $\pm$   1.0 & \phn\phn   1.0 $\pm$   1.0 \nl
$ 23.3 \!-\! 23.6$ & \phn\phn  7.0 $\pm$  2.6 & \phn 2.0   $ \pm$  1.4  & \phn\phn  7.0 $\pm$   2.6 & \phn\phn   3.1 $\pm$   3.0 \nl
$ 23.6 \!-\! 23.9$ & \phn  13.0 $\pm$  3.6    & \phn 1.0   $ \pm$  1.0  &\phn  13.0 $\pm$   3.6 & \phn  11.1 $\pm$   3.8 \nl
$ 23.9 \!-\! 24.2$ & \phn  11.0 $\pm$  3.3    & \phn 1.0   $ \pm$  1.0  &\phn  11.0 $\pm$   3.3 & \phn\phn   9.1 $\pm$ 3.5 \nl
$ 24.2 \!-\! 24.5$ & \phn  22.0 $\pm$  4.7    & \phn 2.0   $ \pm$  1.4  &\phn  22.0 $\pm$   4.7 & \phn  18.2 $\pm$   4.9 \nl
$ 24.5 \!-\! 24.8$ & \phn  36.0 $\pm$  6.0    & \phn 2.0   $ \pm$  1.4  &\phn  36.1 $\pm$   6.0 & \phn  32.2 $\pm$   6.2 \nl
$ 24.8 \!-\! 25.1$ & \phn  53.0 $\pm$  7.3    & \phn  1.0  $ \pm$  1.0  &\phn  53.1 $\pm$   7.3 & \phn  51.2 $\pm$   7.4 \nl
$ 25.1 \!-\! 25.4$ & \phn  81.0 $\pm$  9.0    & \phn  7.0   $ \pm$  2.6 &\phn  81.2 $\pm$   9.0 & \phn  67.7 $\pm$   9.5 \nl
$ 25.4 \!-\! 25.7$ &  115.0 $\pm$ 10.7        &  12.0   $ \pm$ 3.5     & 115.4 $\pm$ 10.8 & \phn  92.2 $\pm$ 11.4 \nl
$ 25.7 \!-\! 26.0$ &  108.0 $\pm$ 10.4        &  17.0  $ \pm$ 4.1      & 108.5 $\pm$ 10.4 & \phn  75.5 $\pm$ 11.3 \nl
$ 26.0 \!-\! 26.3$ &  159.0 $\pm$ 12.6        &  23.0   $ \pm$ 4.8     & 159.9 $\pm$ 12.7 &  115.3 $\pm$ 13.7 \nl
$ 26.3 \!-\! 26.6$ &  229.0 $\pm$ 15.1        &  43.0   $\pm$  6.6     & 230.8 $\pm$ 15.3 &  147.4 $\pm$ 16.8 \nl
$ 26.6 \!-\! 26.9$ &  263.0 $\pm$ 16.2        &  44.0   $ \pm$ 6.6     & 266.2 $\pm$ 16.4 &  180.5 $\pm$ 17.9 \nl
$ 26.9 \!-\! 27.2$ &  320.0 $\pm$ 17.9        &  42.0   $ \pm$ 6.5     & 326.1 $\pm$ 18.2 &  244.1 $\pm$ 19.5 \nl
$ 27.2 \!-\! 27.5$ &  312.0 $\pm$ 17.7        &  66.0   $ \pm$ 8.1     & 324.2 $\pm$ 18.4 &  194.2 $\pm$ 20.4 \nl
$ 27.5 \!-\! 27.8$ &  305.0 $\pm$ 17.5        &  57.0   $ \pm$ 7.5     & 343.4 $\pm$ 19.9 &  229.0 $\pm$ 21.6 \nl
$ 27.8 \!-\! 28.1$ &  292.0 $\pm$ 17.1        &  66.0   $ \pm$ 8.1     & 408.8 $\pm$ 27.5 &  268.0 $\pm$ 29.1 \nl
$ 28.1 \!-\! 28.4$ &  220.0 $\pm$ 14.8        &  81.0   $ \pm$ 9.0     & 436.7 $\pm$ 44.0 &  216.9 $\pm$ 46.1 \nl
$ 28.4 \!-\! 28.7$ &  150.0 $\pm$ 12.2        &  66.0   $ \pm$ 8.1     & 690.7 $\pm$ 87.7 &  291.9 $\pm$ 92.2 \nl
$ 28.7 \!-\! 29.0$ & \phn  97.0 $\pm$ 9.8     &  46.0   $ \pm$ 6.8     & \llap{1}076.5 $\pm$131.4 &  363.7 $\pm$144.4 \nl
$ 29.0 \!-\! 29.3$ & \phn\phn 8.0 $\pm$  2.8  & \phn  3.0 $ \pm$ 1.7   &  159.9 $\pm$ 57.2 &\phn   45.9 $\pm$ 68.4 \nl

\enddata
\end{deluxetable}

\clearpage

\begin{deluxetable}{cccc}
\tablenum{5}
\tablecaption{GCLF $\widetilde{\chi}^2$ Fitting Results  \label{tab5}}
\tablewidth{0pt}
\tablecolumns{3}
\tablehead{
\colhead{$\sigma_V$ } & \colhead{$\widetilde{\chi}^2$ } & 
\colhead{$V_0$ } 
}
\startdata
1.3  &  2.17  &  27.44   \nl
1.4  &  1.62  &  27.59   \nl
1.5  &  1.37  &  27.77   \nl
1.6  &  1.32  &  27.97   \nl
1.7  &  1.42  &  28.21   \nl
1.8  &  1.62  &  28.46   \nl
\enddata
\end{deluxetable}

\clearpage

\begin{deluxetable}{lrll}
\tablenum{6}
\tablecaption{Radial Density Profile \label{tab6}}
\tablewidth{0pt}
\tablecolumns{5}
\tablehead{
\colhead{R (arcsec) } & \colhead{N } & \colhead{A (arcsec$^2$) } & 
\colhead{$\sigma$(n/arcsec$^2$) } 
}
\startdata
$ \phn \phn 0.0 \!-\! 2.0 $& 16 & \phn \phn 12.533&   $1.277  \pm    0.319$  \nl
$ \phn \phn 2.0 \!-\! 4.0 $& 56 & \phn \phn 37.687&   $1.486  \pm    0.199$  \nl
$ \phn \phn 4.0 \!-\! 6.0 $& 86 & \phn \phn 62.845&   $1.368   \pm   0.148$  \nl
$ \phn \phn 6.0 \!-\! 8.0 $& 107& \phn \phn 87.989 &   $1.216  \pm    0.118$ \nl
$ \phn \phn 8.0 \!-\! 10.0$& 113& \phn 113.069&  $0.999  \pm     0.094$ \nl
$ \phn 10.0 \!-\! 15.0 $&  268& \phn 392.731&   $0.682  \pm    0.042$   \nl
$ \phn 15.0 \!-\! 20.0 $&  172& \phn 442.020&   $0.389  \pm    0.030$ \nl
$ \phn 20.0 \!-\! 25.0 $&  98 & \phn 373.326&   $0.263 \pm     0.027$ \nl
$ \phn 25.0 \!-\! 30.0 $&  91 & \phn 459.581&   $0.198 \pm     0.021$ \nl
$ \phn 30.0 \!-\! 35.0 $&  92 & \phn 582.352&   $0.158  \pm    0.017$ \nl
$ \phn 35.0 \!-\! 40.0 $&  68 & \phn 702.702&   $0.097  \pm    0.012$ \nl
$ \phn 40.0 \!-\! 50.0 $&  146&  1762.469&   $0.083  \pm    0.007$ \nl
$ \phn 50.0 \!-\! 60.0 $&  104&  2235.922&   $0.047  \pm    0.005$ \nl
$ \phn 60.0 \!-\! 70.0 $&  90 &  2431.022&  $0.037   \pm   0.004$ \nl
$ \phn 70.0 \!-\! 80.0 $&  88 &  2484.612&  $0.035 \pm     0.004$ \nl
$ \phn 80.0 \!-\! 90.0 $&  81 &  2594.925&  $0.031  \pm    0.004$ \nl
$ \phn 90.0 \!-\! 100.0$&  57 &  2124.154&  $0.027  \pm    0.004$ \nl
$ 100.0 \!-\! 110.0  $&  23 & \phn  898.649&   $0.026  \pm    0.005$   \nl
$ 110.0 \!-\! 130.0  $&  9  & \phn  521.199&   $0.017  \pm    0.006$   \nl
\enddata
\end{deluxetable}

\clearpage

\clearpage

\clearpage

\clearpage

\begin{figure}
\epsscale{1.0}
\plotone{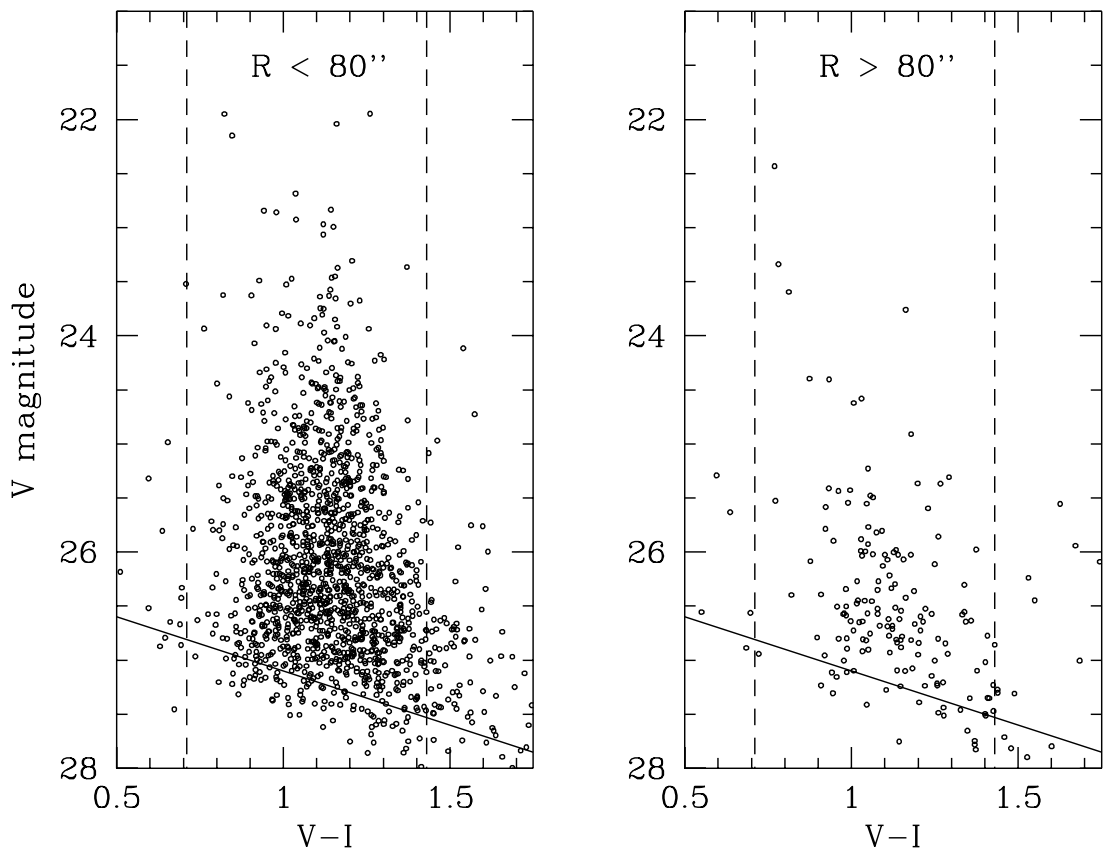}
\caption{
Color-magnitude distribution for all measured starlike objects with
$(V-I)$ colors.  The vast majority of these are globular clusters in
IC 4051.  Objects within $80\arcsec$
of the center of IC 4051 are plotted in the left panel, and
objects lying beyond $80\arcsec$ in the right panel.
The solid line shows the 50\% detection completeness limit in $I$.}
\label{fig4}
\end{figure}
\clearpage

\begin{figure}
\epsscale{1.0}
\plotone{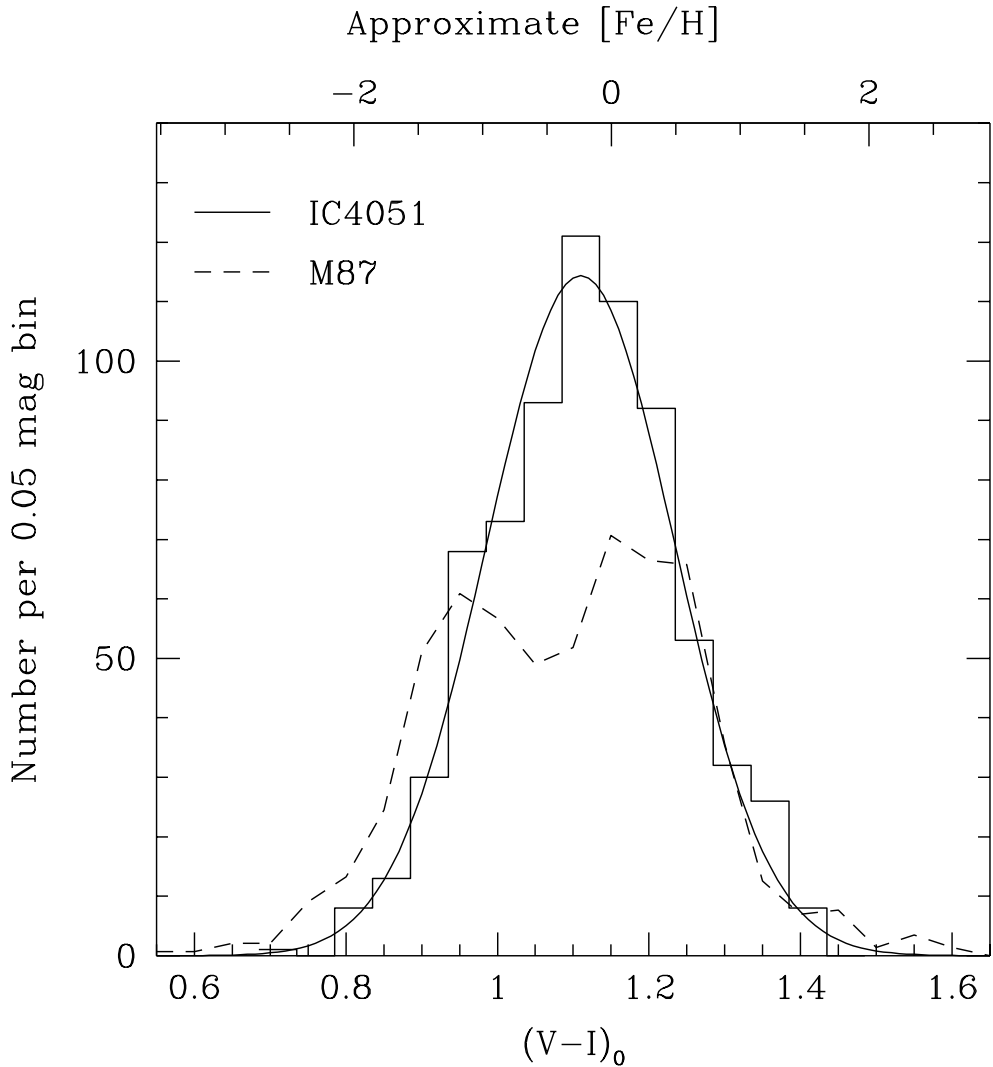}
\caption{
The metallicity distribution function (MDF)
for the globular cluster system in {\ic}.
Number of objects per 0.05-mag bin, after subtraction of
background (see text), is plotted against $(V-I)$.
The [Fe/H] scale at top follows the linear conversion
relation given in the text, and should be taken only
as schematic for [Fe/H] $\gtsim 0$.
The best-fit single Gaussian function is shown, with
$\langle V-I \rangle = 1.12$ and $\sigma(V-I) = 0.13$.
The {\it dashed line} shows the color distribution for the
globular clusters in the Virgo giant M87 (\protect\cite{kun99}).}
\label{fig5}
\end{figure}
\clearpage

\begin{figure}
\epsscale{1.0}
\plotone{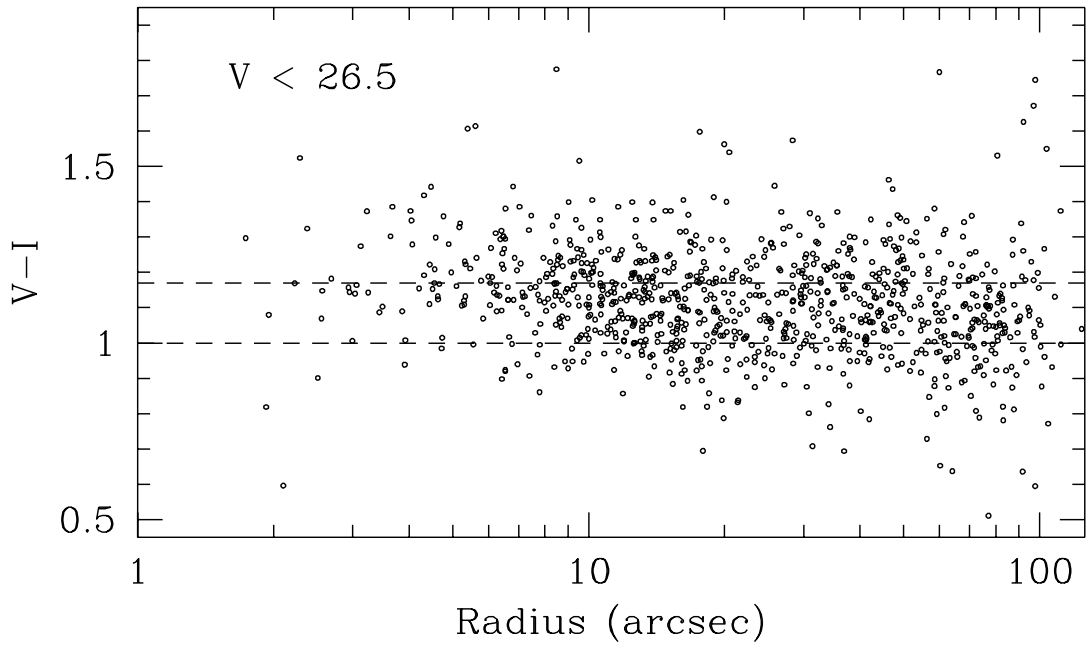}
\caption{
$V-I$ colour plotted against projected galactocentric distance,
for bright objects ($V < 26.0$) for which the photometry is
highly complete at all radii and the measurement uncertainties
are smallest.  Horizontal lines at $(V-I) = 1.17, 1.00$ are drawn
at the suggested two subpopulations; the bluer of the two
components is almost absent within $10''$.}
\label{fig6}
\end{figure}
\clearpage

\begin{figure}
\epsscale{1.0}
\plotone{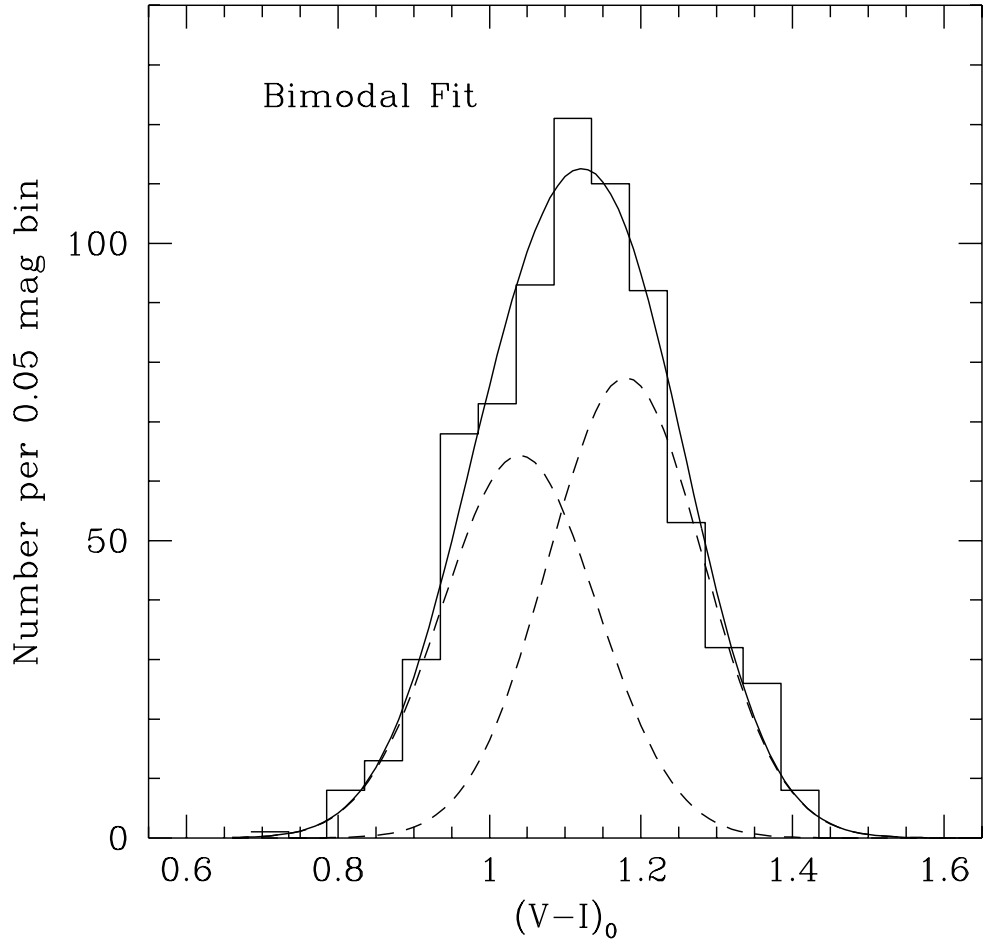}
\caption{
The color distribution of the clusters, fitted by a two-component
Gaussian model as described in the text.  The two subcomponents,
centered at $(V-I) = 1.00, 1.17$ and each with $\sigma = 0.10$,
are shown as the dashed lines, and their sum as the smooth
solid line.}
\label{fig7}
\end{figure}
\clearpage

\begin{figure}
\epsscale{1.0}
\plotone{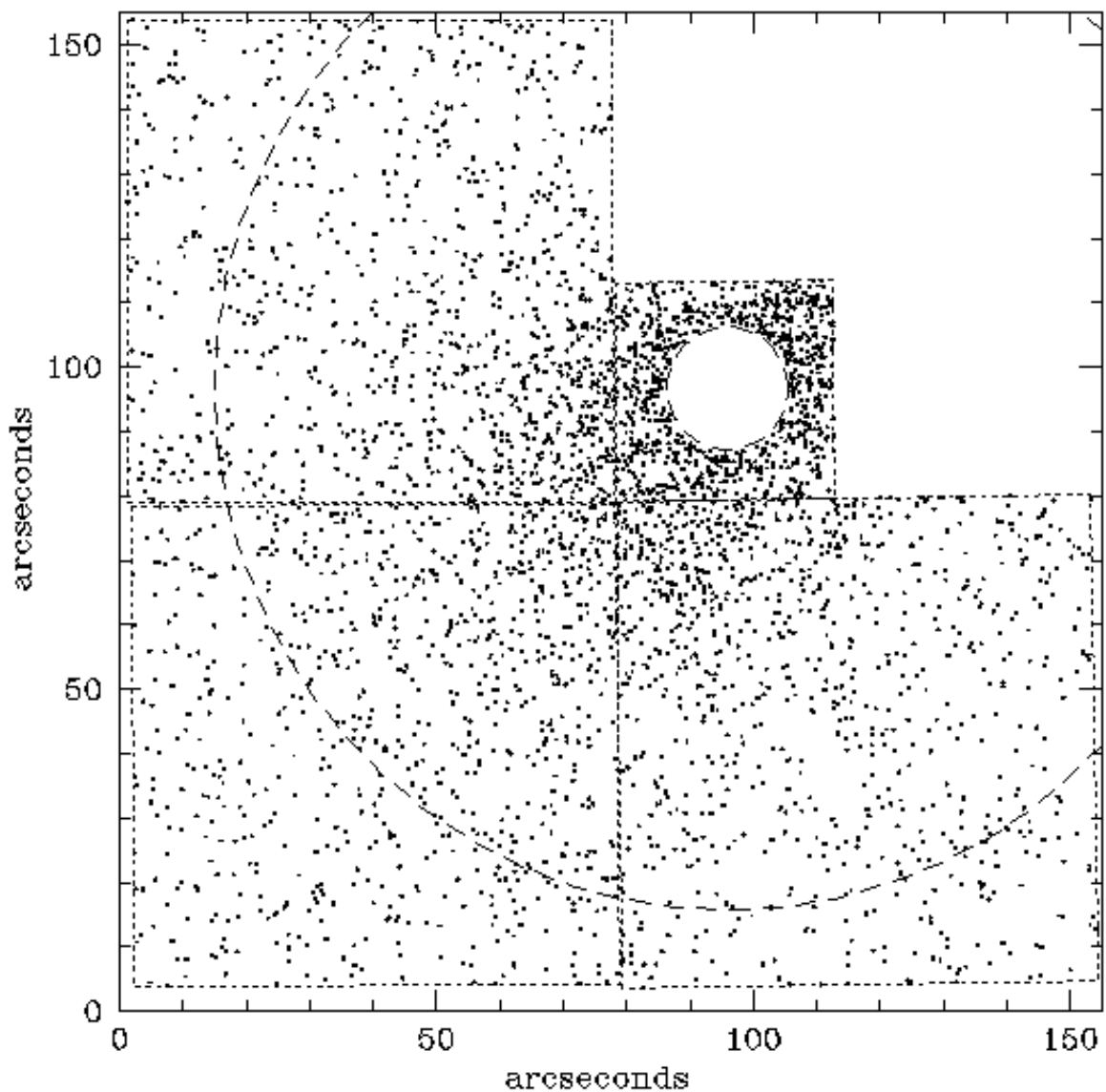}
\caption{
Spatial distribution of the globular clusters
used to define the GCLF. Dotted lines outline the boundaries
of the four CCDs.
The zone outside the outer dashed circle at $R = 80''$ is
used to define the ``background'' number density of objects.
No data within $R = 10''$ (inner dashed circle) were used
for the GCLF analysis.  Note the obvious high 
concentration of the GC population toward the center of IC 4051.}
\label{fig8}
\end{figure}
\clearpage

\begin{figure}
\epsscale{1.0}
\plotone{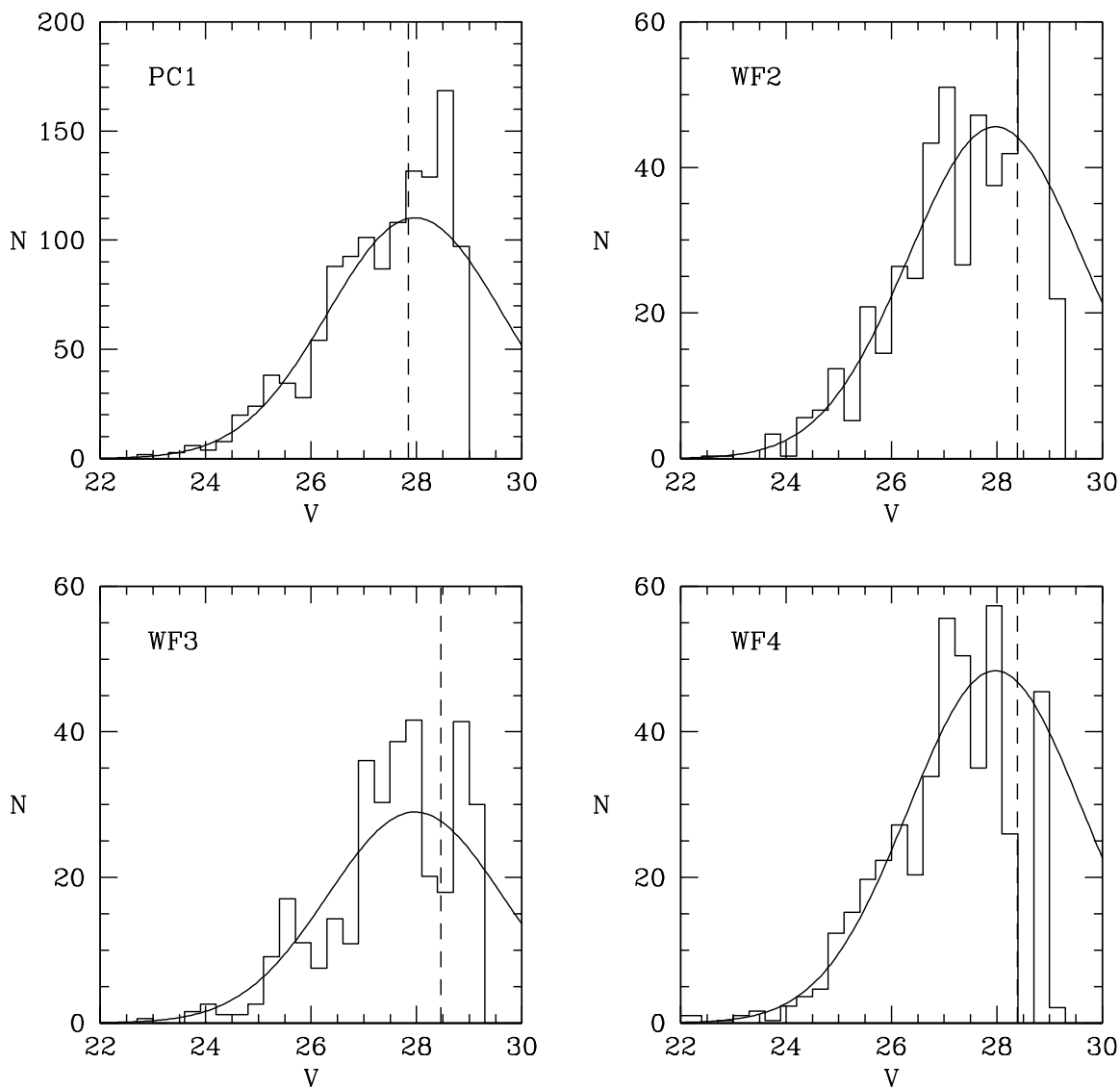}
\caption{
GCLFs (number of objects per 0.3-mag bin, after correction
for incompleteness and subtraction of background)
for the four separate WFPC2 chips.  
The dashed line in each graph represents the 50\% completeness 
limit of the photometry, while the solid line shows the best-fit Gaussian 
function (see next Figure).}
\label{fig9}
\end{figure}
\clearpage

\begin{figure}
\epsscale{1.0}
\plotone{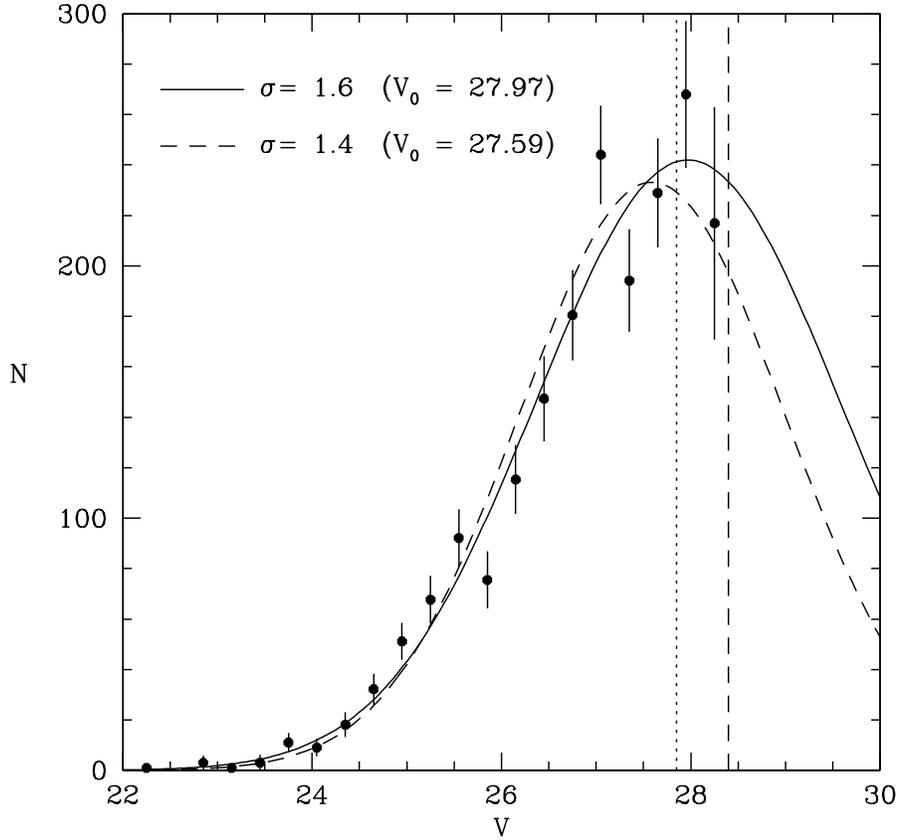}
\caption{
The GCLF for all combined $V$ data (the sum of the four
plots in the previous figure).
The dotted and dashed vertical lines represent 
the $50\%$ completeness 
limit of the PC1 and WF data respectively.
The solid line is the best-fit Gaussian 
function to the binned data for an assumed GCLF dispersion
$\sigma_V = 1.6$, while the dashed line is the best-fit Gaussian
for $\sigma_V = 1.4$.}
\label{fig10}
\end{figure}
\clearpage

\begin{figure}
\epsscale{1.0}
\plotone{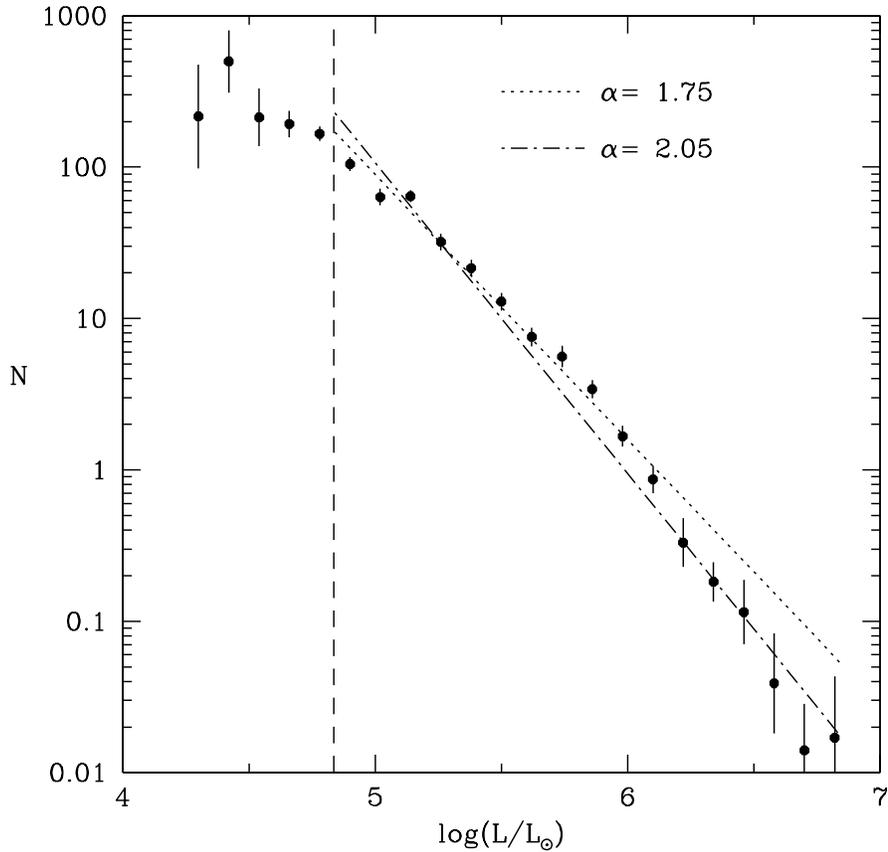}
\caption{
The luminosity distribution function (LDF, or number of clusters
per unit luminosity $L/L_{\odot}$).
The vertical dashed line shows the luminosity at which the
GCLF (from the previous figure) reaches its peak or ``turnover'',
equivalent to log$(L/L_{\odot}) \sim 4.8$.
The dotted line is a weighted best-fit power-law
function $N \sim L^{-\alpha}$ for the restricted
range $4.8 \lesssim \log(L/L_{\odot})
\lesssim 6.5$, yielding $\alpha = 1.75$. 
An unweighted fit to all the data (dot-dashed line)
yields a somewhat steeper slope $\alpha \simeq 2.05$ affected
by the downturn at the bright end.}
\label{fig11}
\end{figure}
\clearpage

\begin{figure}
\epsscale{1.0}
\plotone{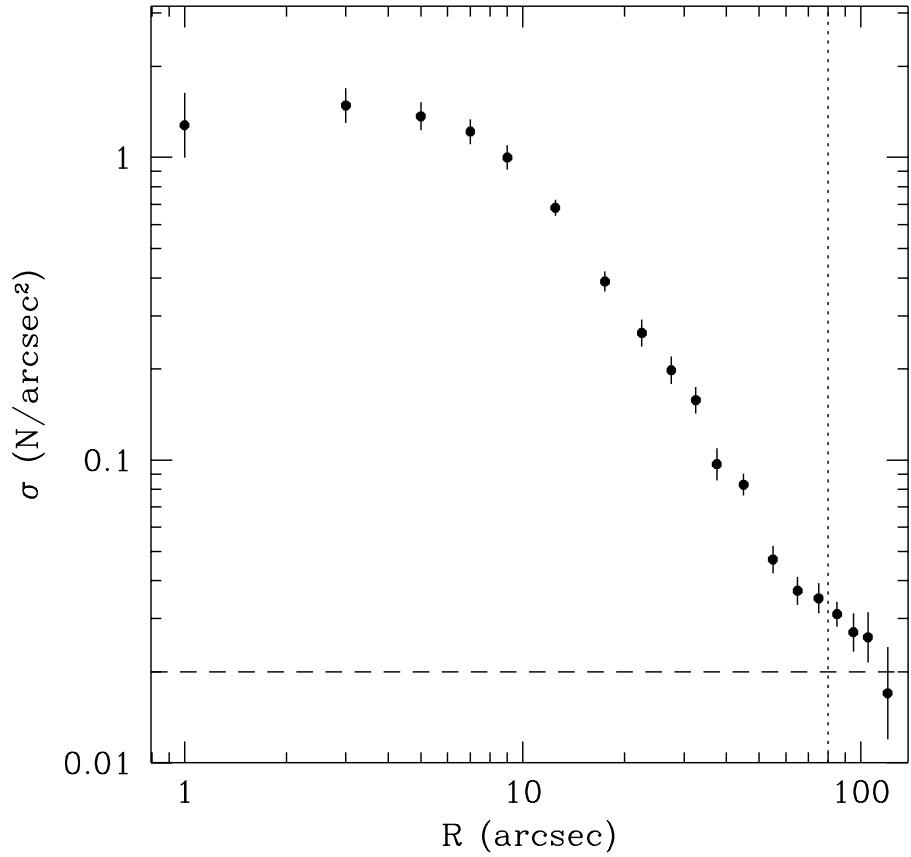}
\caption{
Radial surface density profile for all detected objects with $V \leq 27.0$.   
For $R \gtsim 80''$ (vertical dashed line), the 
number density $\sigma$ begins flattening
off to its asymptotic background level, which we adopt as
$\sigma_b = 0.02$ arcsec$^{-2}$ (horizontal dashed line).}
\label{fig12}
\end{figure}
\clearpage

\begin{figure}
\epsscale{1.0}
\plotone{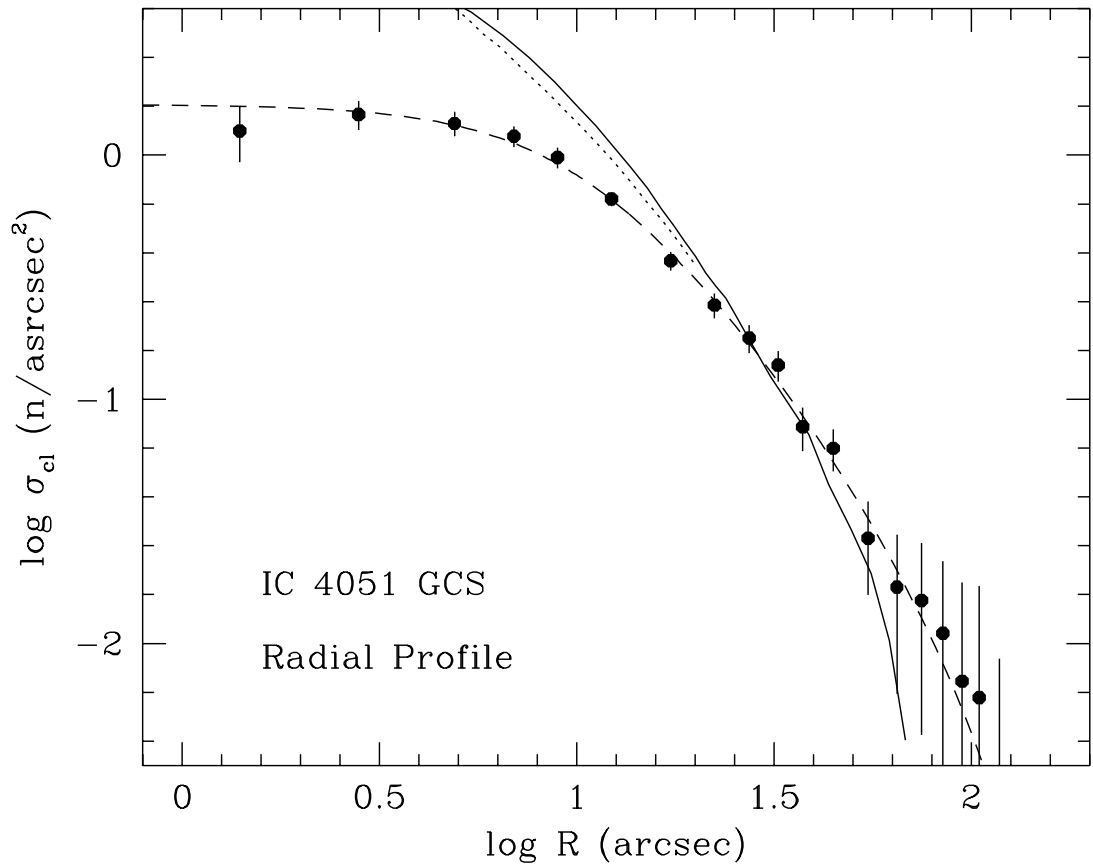}
\caption{
The radial projected density profile of the globular cluster
system.  {\it Solid dots} with error bars are taken from 
Table 7 and assume a background density $\sigma_b = 0.02$ arcsec$^{-2}$.
The solid line is the $R-$band photographic
surface intensity profile of the halo
light, from Strom \& Strom (1978), while the dotted line is the
$R-$band CCD profile from Jorgensen \etal\ (1992).  Both
are shifted vertically arbitrarily to match to the GCS profile.
The dashed line is the best-fit King model discussed in the text,
with core radius $R \simeq 10''$ and $c = 1.45$.}
\label{fig13}
\end{figure}
\clearpage

\end{document}